\documentclass[11pt]{article}

\usepackage[utf8]{inputenc}
\usepackage[T1]{fontenc}
\usepackage{lmodern}
\usepackage[margin=1in]{geometry}
\usepackage{amsmath}
\usepackage{booktabs}
\usepackage{array}
\usepackage{url}
\usepackage[hidelinks]{hyperref}
\usepackage{microtype}
\usepackage{tikz}
\usepackage{pgfplots}
\pgfplotsset{compat=1.16}
\usetikzlibrary{positioning,arrows.meta}

\title{What a Random Draw from the MCP Registry Contains,\\
and What Tool-Use Benchmarks Contain Instead}

\author{Haseeb Mohammed Afsar\\
  \small Independent researcher\\
  \small ORCID: \href{https://orcid.org/0009-0000-4038-1272}{0009-0000-4038-1272}}

\date{August 2026}

\begin{document}
\maketitle

\begin{abstract}
Studies of the Model Context Protocol (MCP) server ecosystem draw their samples in
ways that quietly select for servers that work: reference sets, popularity lists,
hand-curated frames, or pipelines that repair a server until it starts. We report
what an unrepaired probability sample actually contains. From a 24{,}135-server
registry census we draw 400 \texttt{npm}/\texttt{stdio} servers with a published
seed and probe each one over the wire. Only 48.8\% complete an
\texttt{initialize} handshake, against 66.7\% for a hand-curated frame measured
with the same instrument, and the dominant failure is not missing credentials
(13.3\%) but servers that never start at all (37.5\%). Among the 195 that do run,
hard conformance is total: zero fatal JSON~Schema violations across 2{,}766
advertised tools. Optional safety annotations are the real variance, and the
tool-level omission rate on a random draw is 58.8\% against 41.5\% on the curated
frame, so curation flatters this figure too. We then compare the tool
\emph{descriptions} these servers advertise against two tool-use benchmark
corpora using one method held constant. Real MCP tools show 2.8\% near-duplication
at cosine 0.70, and \emph{all} of it lies within single servers: cross-author
near-duplication is 0.0\% at every threshold tested. BFCL~v4 shows 16.7\%, of
which 16.4 points lie \emph{between} independently presented tasks. UltraTool
shows 0.3\%, cleaner than real tools, so this is a property of BFCL and not of
synthetic corpora as a class. Separately, 68.8\% of raw BFCL rows and 85.6\% of
raw UltraTool rows are exact name-plus-description repeats, against 0.4\% for
real MCP, so any statistic computed over these releases without global
deduplication measures repetition rather than tools. All figures regenerate from
released scripts and a published seed.
\end{abstract}

\section{Introduction}

The Model Context Protocol (MCP)~\cite{mcp2024} lets large language model
applications discover and call external tools over a uniform JSON-RPC~2.0
interface. Empirical work on the resulting server population has grown quickly,
and it now includes registry drift measurement~\cite{drift}, internet-facing
security assessment~\cite{exposed}, ecosystem-scale runtime
analysis~\cite{mcpzoo}, and repository-level clone
detection~\cite{cloning}. These studies establish scale. This paper asks a
narrower question that scale alone does not answer.

Every behavioral MCP study must solve the same problem: most published servers
cannot simply be launched and talked to. The solutions in the literature all
involve selection. Reference sets and popularity lists select for servers people
already use. Curated frames select for servers the curator could get working.
MCPZoo~\cite{mcpzoo}, the largest such effort, resolves it by \emph{repairing}
servers with a multi-agent framework until they run, turning 64{,}611 collected
servers into 37{,}288 that support dynamic analysis. Each is a reasonable
engineering answer, and each erases the same quantity: how much of the published
population is simply dead on arrival.

We measure that quantity by not solving the problem. We draw a probability sample
from the registry, probe every draw exactly once with no repair and no
credentials, and record an outcome for all 400, included or excluded with a
reason.

The second half of the paper turns the same instrument outward. Having collected
2{,}766 tool descriptions that real, independently authored, deployed servers
actually advertise, we can ask how they compare with the tool descriptions that
tool-use benchmarks present to models. Existing benchmark criticism has focused
on \emph{evaluators}: Bhat et al.~\cite{benchaudit} find an 18.5\% evaluator-human
misalignment rate across four tool-calling benchmark families including BFCL~v4,
and score spreads of 18.9 points across repeated runs of one setup. We ask a
different question about the same artifacts, concerning the \emph{corpora} being
scored rather than the machinery scoring them.

\paragraph{Contributions.}
\begin{itemize}
  \item \textbf{What an unrepaired random draw contains} (Section~\ref{sec:draw}):
    48.8\% inclusion against 66.7\% curated, with the full failure taxonomy. The
    dominant exclusion is servers that never start, at nearly three times the
    credential-gated rate.
  \item \textbf{What the servers that do run look like}
    (Section~\ref{sec:behavior}): zero fatal schema violations across 2{,}766
    tools, and a safety-annotation omission rate that is 17.3 points worse on a
    random draw than on a curated one.
  \item \textbf{Benchmark corpora against real deployed tools}
    (Section~\ref{sec:redundancy}): one similarity method held constant across
    three corpora, with the within-unit and cross-unit decomposition that
    separates a project repeating itself from a benchmark repeating itself.
  \item \textbf{A released, seeded, re-runnable pipeline} (Section~\ref{sec:data}).
    Every number here regenerates from a committed script, a committed seed, and
    a hash-pinned frame.
\end{itemize}

\section{Related work}
\label{sec:related}

\paragraph{Registry measurement.}
Bharti~\cite{drift} tracks 120 snapshots of the MCP registry over 88.6 days across
19{,}099 servers, and reports that only 8.6\% of servers ever rewrite their
descriptions while the most active 5\% generate 61\% of change events. That work
supersedes single-snapshot registry audits on the question of \emph{drift}. Our
census tier is two snapshots and we make no drift claim; we use it only to size
the population and to construct a sampling frame.

\paragraph{Behavioral and security measurement.}
Chen et al.~\cite{mcpzoo} build MCPZoo, the largest runtime collection to date,
and use it to show that existing MCP security scanners are unreliable: fewer than
half of sampled alerts survive manual validation. Padilla~\cite{exposed}
dynamically audits 414 internet-facing MCP servers with a purpose-built
framework, finding 68 reportable vulnerabilities and reporting that 41.6\% of
confirmed servers disappear within three days between measurement runs. Our
behavioral sample is two orders of magnitude smaller than MCPZoo's and we make no
security claim. The complementarity is specific: MCPZoo's repair pipeline is
designed to convert non-starting servers into starting ones, and our measurement
is of exactly the population that pipeline is built to rescue. Padilla's
three-day disappearance rate is independent corroboration, on the remote tier,
of the churn our 37.5\% non-start rate implies on the local tier.

\paragraph{Duplication.}
Kim et al.~\cite{cloning} measure repository-level cloning across 7{,}508 MCP
repositories and 87{,}564 tools using lexical and fuzzy-structural similarity,
manually verifying 60\% of high-Jaccard and 85\% of high-ssdeep candidates as true
clones. Their unit is source code; ours is the advertised tool interface. Their
result and ours are compatible and the tension between them is informative; we
treat it as a threat to validity and test it directly in
Section~\ref{sec:threats}.

\paragraph{Benchmark validity.}
Bhat et al.~\cite{benchaudit} audit the evaluators of four tool-calling benchmark
families. To our knowledge no published work measures duplication \emph{within}
these corpora or compares their tool distributions against tools that are
actually deployed.

\section{Method}
\label{sec:method}

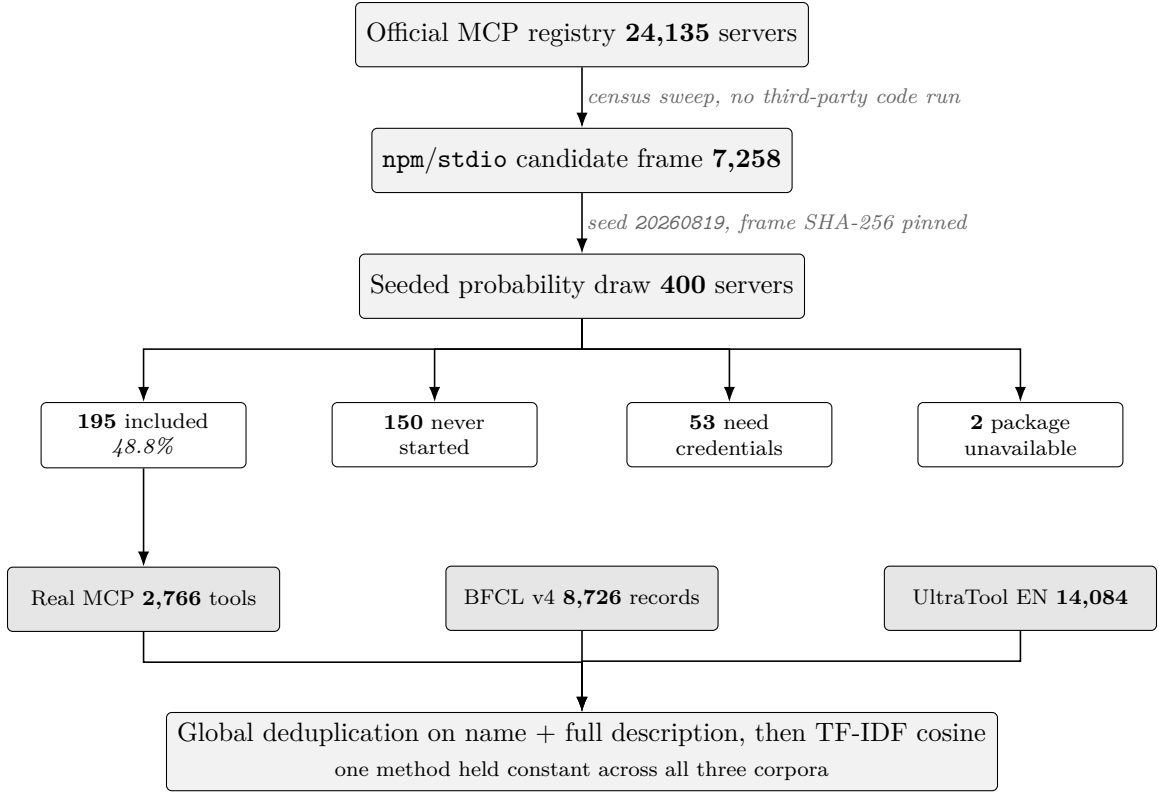
\begin{figure}[t]
\centering
\begin{tikzpicture}[
  font=\small,
  stg/.style={draw, rounded corners=2pt, fill=black!5, align=center, inner sep=4pt,
              minimum height=8.5mm, minimum width=52mm},
  res/.style={draw, rounded corners=2pt, fill=white, align=center, inner xsep=3pt, inner ysep=4pt,
              minimum height=8.5mm, text width=25mm, font=\scriptsize},
  cor/.style={draw, rounded corners=2pt, fill=black!10, align=center, inner sep=4pt,
              minimum height=8.5mm, minimum width=36mm, font=\scriptsize},
  ar/.style={-{Latex[length=2mm]}, semithick},
  nt/.style={font=\scriptsize\itshape, text=black!60, align=left, inner sep=1.5pt},
]
\node[stg] (reg) {Official MCP registry \textbf{24{,}135} servers};
\node[stg, below=8mm of reg] (frm) {\texttt{npm}/\texttt{stdio} candidate frame \textbf{7{,}258}};
\node[stg, below=8mm of frm] (drw) {Seeded probability draw \textbf{400} servers};
\draw[ar] (reg) -- node[right, nt] {census sweep, no third-party code run} (frm);
\draw[ar] (frm) -- node[right, nt] {seed \texttt{20260819}, frame SHA-256 pinned} (drw);
\node[res, below=11mm of drw, xshift=-58mm] (inc) {\textbf{195} included\\\emph{48.8\%}};
\node[res, below=11mm of drw, xshift=-19.5mm] (hsf) {\textbf{150} never\\started};
\node[res, below=11mm of drw, xshift=19.5mm] (crd) {\textbf{53} need\\credentials};
\node[res, below=11mm of drw, xshift=58mm] (unv) {\textbf{2} package\\unavailable};
\foreach \n in {inc,hsf,crd,unv} \draw[ar] (drw.south) -- ++(0,-4mm) -| (\n.north);
\node[cor, below=13mm of inc] (tls) {Real MCP \textbf{2{,}766} tools};
\node[cor, below=13mm of crd, xshift=-19.5mm] (bf) {BFCL v4 \textbf{8{,}726} records};
\node[cor, below=13mm of unv] (ul) {UltraTool EN \textbf{14{,}084}};
\draw[ar] (inc) -- (tls);
\node[stg, below=10mm of drw.south, yshift=-42mm, minimum width=100mm] (red)
  {Global deduplication on name $+$ full description, then TF-IDF cosine\\[1pt]
   \scriptsize one method held constant across all three corpora};
\foreach \n in {tls,bf,ul} \draw[ar] (\n.south) -- ++(0,-4mm) -| (red.north);
\end{tikzpicture}
\caption{The measurement pipeline. The census tier executes no third-party code and
supplies the sampling frame; the dynamic tier probes each drawn server exactly once,
with no repair, no credentials and no retry, and records an outcome for all 400. The
selection points that other MCP studies resolve by curation or repair are the two
labelled arrows and the four-way split beneath them.}
\label{fig:pipeline}
\end{figure}

Figure~\ref{fig:pipeline} shows the whole pipeline and the quantity at each stage.

\subsection{Census tier}
A harvester sweeps the entire official MCP registry
(\url{https://registry.modelcontextprotocol.io/v0/servers}) by cursor pagination
and records only self-declared metadata: deployment model, package ecosystem,
declared transport, lifecycle status, and pinned \texttt{\$schema} revision. No
third-party code runs, so this tier scales to the whole published population. The
sweep records page count, version-row count, and a \texttt{sweepComplete} flag, so
a truncated run is marked a lower bound rather than reported silently.

We report two snapshots, 2026-07-14 and 2026-08-22. Both are complete sweeps. A
registry sweep cannot be reconstructed after the fact, because the population
moves; each snapshot is therefore released as an aggregate at the moment it was
taken.

\subsection{Sampling}
The frame is the \texttt{npm}-published, \texttt{stdio}-declared, active servers
the census emits, 7{,}258 candidates at the 2026-08-22 snapshot. We draw
$n=400$ without replacement: canonical sort by registry identifier, then partial
Fisher--Yates driven by a seeded \texttt{mulberry32} generator, seed
\texttt{20260819}. The draw manifest records the SHA-256 of the exact frame bytes,
so a redraw against a moved frame is detected rather than assumed equivalent.

\subsection{Probing}
Each drawn package is launched over stdio via \texttt{npx} and probed once with
\texttt{mcp-probe}, which speaks the MCP wire protocol directly, performs the
\texttt{initialize} handshake, enumerates tools via \texttt{tools/list}, and
validates each tool's JSON~Schema~\cite{jsonschema} against the constraints the
specification places on tool definitions. No credentials are ever supplied, no
side-effecting tool is invoked, no server is repaired or retried, and every draw
is recorded with an outcome.

\subsection{Redundancy measurement}
Corpora are compared on one method held constant. The unit is tool name
concatenated with full description. Every corpus is \textbf{globally
deduplicated} on that exact key before any similarity is computed; this is
enforced in code rather than left to discipline, because a per-file deduplication
key in an earlier iteration of this work missed cross-file repeats and inflated a
measured figure by fourteen points. Vectorisation is TF-IDF over word unigrams and
bigrams with sublinear term frequency, fit separately per corpus, never on a
shared vocabulary. Similarity is cosine. \emph{Redundancy at threshold $t$} is the
share of deduplicated tools having at least one \emph{other} tool at cosine
$\geq t$; we state the definition because a different one yields a different
number.

We also report the exact-duplicate rate of each \emph{raw} release, which is a
distinct and more consequential quantity than the near-duplicate rate.

\subsection{Corpora}
\emph{Real MCP}: the 2{,}766 tools advertised by the 195 included servers.
\emph{BFCL~v4}~\cite{bfcl}: every tool definition in every released row's
\texttt{function} list, 8{,}726 records. Seven of the twenty released files expose no
\texttt{function} list at all, including the four multi-turn files and the memory
file, whose tool definitions live separately; those contribute nothing here and
the multi-turn tool documents are out of scope. \emph{UltraTool}~\cite{ultratool}:
every tool in every English-split row's \texttt{tools} list, development and test
together, 14{,}084 records. Every source file is pinned by SHA-256 in the released
provenance record.

\section{RQ1: what an unrepaired random draw contains}
\label{sec:draw}

Table~\ref{tab:inclusion} gives the outcome of all 400 draws.

\begin{table}[h]
\centering
\begin{tabular}{lrr}
\toprule
Outcome & Count & Share \\
\midrule
Included (handshake completed)      & 195 & 48.8\% \\
Excluded: handshake failed          & 150 & 37.5\% \\
Excluded: needs credentials         & 53  & 13.3\% \\
Excluded: package unavailable       & 2   & 0.5\%  \\
\bottomrule
\end{tabular}
\caption{Outcome of every draw. No server was dropped silently.}
\label{tab:inclusion}
\end{table}

\begin{figure}[t]
\centering
\begin{tikzpicture}
\begin{axis}[
  width=0.78\linewidth, height=4.6cm, font=\small,
  ybar, bar width=20pt, enlarge x limits=0.55,
  symbolic x coords={inclusion,omission}, xtick=data,
  xticklabels={servers that start,tools missing annotations},
  ymin=0, ymax=82, ytick={0,20,40,60}, ylabel={percent},
  grid=major, grid style={black!10}, tick align=outside,
  nodes near coords, nodes near coords style={font=\scriptsize},
  point meta=explicit symbolic,
  legend style={font=\scriptsize, at={(0.5,1.04)}, anchor=south, legend columns=2,
                draw=black!25, /tikz/every even column/.append style={column sep=8pt}},
  legend image code/.code={\draw[#1] (0cm,-0.09cm) rectangle (0.34cm,0.13cm);},
]
\addplot[draw=black, fill=black!45] coordinates {(inclusion,66.7)[66.7] (omission,41.5)[41.5]};
\addlegendentry{hand-curated frame ($n=24$)}
\addplot[draw=black, fill=black!12] coordinates {(inclusion,48.8)[48.8] (omission,58.8)[58.8]};
\addlegendentry{random draw ($n=400$)}
\end{axis}
\end{tikzpicture}
\caption{Curation flatters both headline numbers, and in opposite directions. The same
instrument reports a higher start rate and a lower annotation-omission rate on a
hand-curated frame than on a probability sample of the same population.}
\label{fig:curation}
\end{figure}
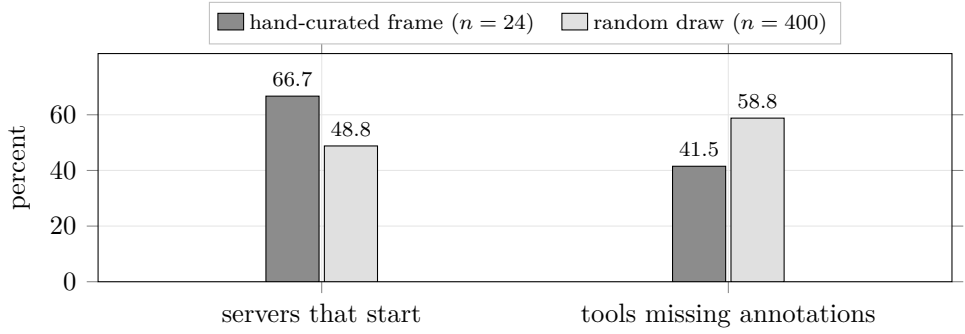

Two things follow.

First, \textbf{a random draw includes far less than a curated one}. The same
instrument, run against a 24-server hand-curated frame of reference and popular
community servers, included 16 of 24, or 66.7\%. The 17.9-point gap is the size of
the selection effect that curation introduces, measured rather than asserted.

Second, \textbf{the dominant failure mode is not the one the literature
anticipates}. Credential gating is the standard explanation for why a public MCP
server cannot be probed, and it accounts for 53 of 400 draws. Servers that simply
do not start account for 150, nearly three times as many. On the frame that
behavioral MCP studies actually sample from, roughly two in five published entries
are inert.

Figure~\ref{fig:curation} puts the two curation effects side by side. They run in
opposite directions, which is what makes curation hard to correct for after the fact:
it raises the apparent health of the population on one axis and lowers it on the other.

This is the quantity that repair-based pipelines are built to eliminate. That is a
sound engineering choice for their purposes and it is precisely why their samples
cannot report this number.

\section{RQ2: what the servers that do run look like}
\label{sec:behavior}

The 195 included servers advertise 2{,}766 tools; 2{,}759 carry a description.
Tool counts per server are highly skewed: minimum 1, median 8, 95th percentile 46,
maximum 300. One included server advertises no tools at all.

\subsection{Hard conformance}
\textbf{Zero of 2{,}766 tools carry a fatal JSON~Schema violation}, and zero of
195 servers have any. No missing schema, no invalid \texttt{type}, no malformed
\texttt{properties} or \texttt{required}. This replicates a 200-tool result from
the earlier curated frame at roughly fourteen times the scale and on a random
rather than selected sample, which makes it a structural property rather than a
small-sample artifact. The intuition that MCP tool schemas are frequently
malformed is not supported.

\subsection{Safety annotations}
Optional \texttt{annotations} (\texttt{readOnlyHint}, \texttt{destructiveHint},
\texttt{idempotentHint}, \texttt{openWorldHint}) tell an agent whether a tool is
safe to call \emph{before} calling it. Of the 194 included servers advertising at
least one tool, 72 annotate every tool and 122 annotate none. \textbf{1{,}626 of
2{,}766 tools (58.8\%) carry no annotations.}

The curated frame gave 41.5\%. Curation therefore flatters this figure by 17.3
points, for the same reason it flatters inclusion: reference servers annotate, and
curated frames are full of reference servers. The lower number should not be cited
as an ecosystem rate.

The split is close to bimodal at the server level: 194 of 194 servers in this
sample are all-or-nothing, with no partial server observed. We deliberately do not
state this as an absolute. A single sample observing none of a rare category
supports an upper bound, not a denial: given 0 of 194, the one-sided 95\% upper
bound on the prevalence of partial annotation is 1.53\%. An earlier unreleased run
of this study reported four partial servers at $n=214$; the present run does not
reproduce that and we report the disagreement without explaining it.

\subsection{Protocol versions}
Four versions were negotiated across 195 servers: \texttt{2025-06-18} on 192,
and \texttt{2024-11-05}, \texttt{2025-03-26} and \texttt{2025-11-25} on one each.
The last of these is \emph{newer} than the baseline our client advertises, so
version spread in the deployed population runs in both directions and not only as
a lagging tail.

\section{RQ3: benchmark corpora against real deployed tools}
\label{sec:redundancy}

\subsection{Raw releases are mostly repetition}
Table~\ref{tab:exact} reports exact name-plus-description duplicates in each raw
release.

\begin{table}[h]
\centering
\begin{tabular}{lrrr}
\toprule
Corpus & Raw records & Exact duplicates & Share \\
\midrule
Real MCP (195 servers)   & 2{,}766  & 10     & 0.4\%  \\
BFCL v4                  & 8{,}726  & 6{,}002  & 68.8\% \\
UltraTool EN (dev+test)  & 14{,}084 & 12{,}052 & 85.6\% \\
\bottomrule
\end{tabular}
\caption{Exact-duplicate contamination of the raw releases.}
\label{tab:exact}
\end{table}

Any statistic computed over these files without global deduplication measures how
often a benchmark repeats a task, not how many tools it contains. We note that
this percentage is sensitive to which files are included, since a benchmark
reusing one tool across many rows produces a high rate by construction; the
post-deduplication rates below are considerably more stable and the two should be
read together.

\subsection{Near-duplication, and where it lives}
Table~\ref{tab:redundancy} gives the redundancy rate after global deduplication,
decomposed by whether the near-duplicate partner lies inside the same authoring
unit or outside it. For real MCP the unit is the server; for the benchmarks it is
the task row.

\begin{table}[h]
\centering
\begin{tabular}{llrrrr}
\toprule
Corpus (deduplicated $n$) & Pairs counted & 0.70 & 0.80 & 0.85 & 0.90 \\
\midrule
Real MCP (2{,}756)     & all           & 2.8\%  & 1.1\% & 0.5\% & 0.0\% \\
                       & cross-server  & \textbf{0.0\%}  & \textbf{0.0\%} & \textbf{0.0\%} & \textbf{0.0\%} \\
\addlinespace
BFCL v4 (2{,}724)      & all           & 16.7\% & 9.7\% & 6.2\% & 2.7\% \\
                       & cross-task    & \textbf{16.4\%} & \textbf{9.6\%} & \textbf{6.2\%} & \textbf{2.7\%} \\
\addlinespace
UltraTool EN (2{,}032) & all           & 0.3\%  & 0.0\% & 0.0\% & 0.0\% \\
                       & cross-task    & 0.3\%  & 0.0\% & 0.0\% & 0.0\% \\
\bottomrule
\end{tabular}
\caption{Redundancy after global deduplication, by cosine threshold.}
\label{tab:redundancy}
\end{table}

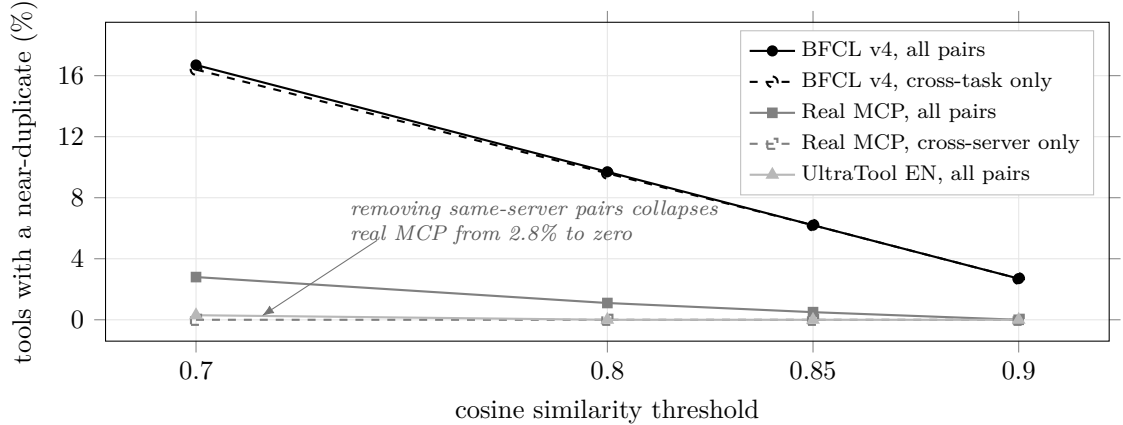
\begin{figure}[t]
\centering
\begin{tikzpicture}
\begin{axis}[
  width=0.9\linewidth, height=5.8cm, font=\small,
  xlabel={cosine similarity threshold}, ylabel={tools with a near-duplicate (\%)},
  xmin=0.678, xmax=0.922, ymin=-1.4, ymax=19.5,
  xtick={0.70,0.80,0.85,0.90}, ytick={0,4,8,12,16},
  grid=major, grid style={black!10}, tick align=outside,
  legend style={font=\scriptsize, at={(0.985,0.975)}, anchor=north east, draw=black!25,
                row sep=0.5pt},
  legend cell align=left,
]
\addplot[thick, black, mark=*, mark size=1.7pt] coordinates {(0.70,16.7)(0.80,9.7)(0.85,6.2)(0.90,2.7)};
\addlegendentry{BFCL v4, all pairs}
\addplot[thick, black, dashed, mark=o, mark size=2pt] coordinates {(0.70,16.4)(0.80,9.6)(0.85,6.2)(0.90,2.7)};
\addlegendentry{BFCL v4, cross-task only}
\addplot[thick, black!50, mark=square*, mark size=1.6pt] coordinates {(0.70,2.8)(0.80,1.1)(0.85,0.5)(0.90,0.0)};
\addlegendentry{Real MCP, all pairs}
\addplot[thick, black!50, dashed, mark=square, mark size=1.9pt] coordinates {(0.70,0.0)(0.80,0.0)(0.85,0.0)(0.90,0.0)};
\addlegendentry{Real MCP, cross-server only}
\addplot[thick, black!28, mark=triangle*, mark size=2.2pt] coordinates {(0.70,0.3)(0.80,0.0)(0.85,0.0)(0.90,0.0)};
\addlegendentry{UltraTool EN, all pairs}
\node[font=\scriptsize\itshape, text=black!65, anchor=west, align=left] (ann) at (axis cs:0.735,6.4)
  {removing same-server pairs collapses\\real MCP from 2.8\% to zero};
\draw[-{Latex[length=1.6mm]}, black!55] (axis cs:0.744,5.2) -- (axis cs:0.716,0.25);
\end{axis}
\end{tikzpicture}
\caption{Near-duplication after global deduplication. Solid lines count every pair;
dashed lines count only pairs spanning different authoring units. BFCL's redundancy
survives the restriction almost unchanged, so it lies between independently presented
tasks. Real MCP's does not survive it at all. The Real MCP cross-server series lies on
zero at every threshold.}
\label{fig:redundancy}
\end{figure}

The decomposition changes the finding. \textbf{Every near-duplicate among real
MCP tools lies inside a single server}: the \texttt{list\_x} / \texttt{get\_x} /
\texttt{create\_x} families that one project naturally produces. Across
independent authors, near-duplication is 0.0\% at every threshold tested.
\textbf{BFCL's redundancy is the opposite kind}: 16.4 of its 16.7 points lie
between independently presented tasks, and it is the only corpus of the three with
near-duplicates surviving at cosine 0.90.

So the claim is not that BFCL is roughly six times more redundant than real
tools. It is that \emph{BFCL repeats itself across tasks, and real MCP does not
repeat itself across authors.}

\textbf{UltraTool is cleaner than real deployed tools}, at 0.3\% against 2.8\%.
Two synthetic corpora built for the same purpose give opposite answers, so no
claim about synthetic tool corpora as a class is supported by this evidence, and
we make none.

\section{Ecosystem context}
\label{sec:context}

The two census snapshots frame the above. The population grew from 16{,}548 to
24{,}135 unique servers between 2026-07-14 and 2026-08-22, about 195 net new
servers per day, and both sweeps completed.

\begin{table}[h]
\centering
\begin{tabular}{lrrr}
\toprule
Deployment model & 2026-07-14 & 2026-08-22 & Change \\
\midrule
Package-only (installed locally) & 8{,}340 (50.4\%)  & 10{,}530 (43.6\%) & $-6.8$pp \\
Remote-only (hosted HTTP/SSE)    & 7{,}057 (42.6\%)  & 12{,}004 (49.7\%) & $+7.1$pp \\
Both                             & 852 (5.1\%)     & 1{,}224 (5.1\%)   & $-0.1$pp \\
Neither declared                 & 299 (1.8\%)     & 377 (1.6\%)     & $-0.2$pp \\
\bottomrule
\end{tabular}
\caption{Deployment model at two complete sweeps 39 days apart.}
\label{tab:deploy}
\end{table}

Remote-only overtook package-only in this window, growing 70.1\% against 26.3\%.
Two snapshots cannot establish a trend and we claim none; we report a change
between two measured endpoints.

The consequence that matters for this paper is methodological. The
\texttt{npm}/\texttt{stdio} slice, which is what a local behavioral instrument can
reach, grew in absolute terms from 5{,}804 to 7{,}414 servers but \emph{fell} as a
share of the population, from 35.1\% to 30.7\%. A stdio-only instrument therefore
covers a shrinking minority of the ecosystem, and this applies to our behavioral
tier as much as to anyone else's.

\section{Threats to validity}
\label{sec:threats}

\paragraph{The handshake filter, tested directly.}
Kim et al.~\cite{cloning} find pervasive code cloning in the MCP repository
population. Our cross-author redundancy of 0.0\% would be flattering rather than
informative if our probe systematically discarded clones, since we observe only
the 48.8\% of draws that start. We tested this three ways. First, using the
npm-authored package description, which exists for started and non-started servers
alike, both groups show 0.0\% near-duplication at every threshold; this test has
low power, since it reads descriptions rather than code and both groups sit at
zero. Second, author-family concentration by npm scope is comparable across the
two groups, with the largest family in the sample (7 packages) entirely included
and the next (8 packages) entirely excluded. Third, and most directly, the author
shipping the most servers that all started contributes 125 tools across 7 servers
with a maximum \emph{cross-server} cosine of 0.623, below the lowest threshold
reported. We find no evidence that the filter selects against clones. We cannot
exclude the possibility that clones copy implementations while rewriting tool
descriptions, which our method would miss and theirs would catch; if that is what
is happening, the gap between code-level and interface-level duplication is itself
a result worth reporting.

\paragraph{The cross-unit columns are not power-matched.}
After deduplication BFCL averages 1.8 tools per task row against real MCP's 14.2
per server, so removing same-unit pairs subtracts far less from BFCL by
construction. The BFCL cross-task figure sitting near its all-pairs figure is
partly an artifact of this. The real MCP collapse from 2.8\% to 0.0\% runs in the
opposite direction and is not explained by it.

\paragraph{Corpus rate and large servers.}
Capping the real MCP corpus at 10, 25 and 50 tools per server yields 0.9\%, 2.2\%
and 3.0\% at threshold 0.70, bracketing the uncapped 2.8\%. The rate is not driven
by the two servers advertising 300 and 122 tools.

\paragraph{Scope of the behavioral tier.}
\texttt{npm}/\texttt{stdio} only, which is 30.7\% of the population and falling.
Remote servers, PyPI and OCI packages are out of frame. Inclusion is decided by a
single probe attempt with no retry, so transient failures are counted as
exclusions and 48.8\% is a lower bound on the fraction that could ever start.

\paragraph{Similarity is lexical.}
TF-IDF cosine measures lexical overlap, not semantic equivalence. Two tools doing
the same thing in different words are counted as distinct in every corpus. This
biases all three rates downward and we have no reason to think it biases them
unequally, but we have not shown that.

\paragraph{Census.}
Registry metadata is self-declared; a server misdeclaring its transport is counted
as declared. Two snapshots support a difference, not a trend. A GitHub topic count
reported in an earlier version of this work as ecosystem context was unavailable
on the second sweep and is not used in any claim.

\paragraph{Timing.}
Wall-clock times were collected but include first-run \texttt{npx} install cost
and were measured under concurrency with a kill timeout. They are not a latency
measurement and none is reported.

\section{Data and code availability}
\label{sec:data}

The \texttt{mcp-probe} instrument, the harvester, the seeded draw script, the
resumable probe runner, the aggregation script, the redundancy measurement with
its self-test, and the threat-test suite are open source at
\url{https://github.com/itguruhaseeb/mcp-probe}. The tool release and dataset are
archived on Zenodo under the concept DOI
\href{https://doi.org/10.5281/zenodo.21347997}{10.5281/zenodo.21347997}, which
always resolves to the latest version~\cite{mcpprobe_dataset}.

Every figure in this paper regenerates from a committed script plus the published
seed \texttt{20260819} and the frame hash recorded in the draw manifest. The
per-server outcome of all 400 draws is released, so the aggregates can be
independently recounted rather than trusted.

\section*{Acknowledgements}
We thank the maintainers of the Model Context Protocol specification and its
reference server implementations, against which \texttt{mcp-probe} is tested.

\end{document}